\documentclass[acus]{JAC2003}

\usepackage{graphicx}
\usepackage{booktabs}

\usepackage{pdflscape}

\begin{document}
\title{Closed Loop Testing of Microphonics Algorithms Using a Cavity Emulator
\thanks{The authors of this work grant the arXiv.org and LLRF Workshop's International Organizing Committee a non-exclusive and irrevocable license to distribute the article, and certify that they have the right to grant this license.}\thanks{ Work supported by Fermi Research Alliance LLC. Under DE-AC02-07CH11359 with U.S. DOE.}}

\author{S. Raman\thanks{ ssankar@fnal.gov}, P. Varghese, B. Chase, S. Ahmed, C. Fulz, P. Hanlet, D. Klepec \\
Fermi National Accelerator Laboratory (FNAL), Batavia, IL 60510, USA}

\maketitle

\begin{abstract}
  An analog crystal filter based cavity emulator is modified with reverse biased varactor diodes to provide a tuning range of around 160 Hz. The piezo drive voltage of the resonance controller is used to detune the cavity through the bias voltage. A signal conditioning and summing circuit allows the introduction of microphonics disturbance from a signal source or using real microphonics data from cavity testing. This setup is used in closed loop with a cavity controller and resonance controller to study the effectiveness of resonance control algorithms suitable for superconducting cavities.
\end{abstract}

\section{INTRODUCTION}
Low-level Radio Frequency (LLRF) systems are a critical subsystem of particle accelerators providing precision control of RF cavity field amplitude and phase. Cavity emulators provide a means of testing LLRF control hardware and firmware operation without waiting for the RF systems to being installed, which happens in the later stages of a project. An analog cavity emulator based on a crystal filter was developed for testing the various RF systems of the Fermilab PIP-II superconducting linac project[1,3]. This design is modified to make the crystal filter resonant frequency tunable by using reverse biased varactor diodes. The resulting narrow tuning range of about 160 Hz makes it suitable to study the effects of microphonics disturbances on the cavity resonant frequency as well as the closed loop feedback algorithms to suppress these disturbances. Detuning data from cavity operation can be played through an arbitrary waveform generator to provide the microphonics disturbance input and check how the various modes are suppressed by the feedback algorithms.
\par
A voltage summing circuit for the tuning bias allows multiple disturbance sources to be included along with the resonance controller output to provide the feedback correction.
 Modifications were also made to the previous design to allow the emulator to be used for all three RF frequencies of the PIP-II linac, 162.5 MHz, 325 MHz and 650 MHz. The emulator design and preliminary 
measurements are presented in this paper.
\begin{figure}[!t]
\centering
\includegraphics[width=3.0in]{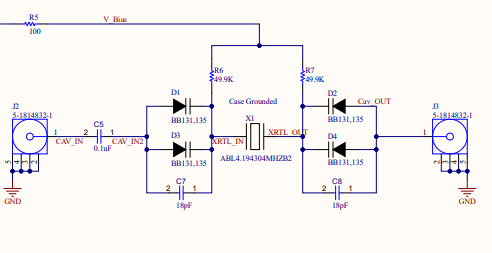}
\caption{Crystal Filter with Varactor Diodes}
\label {fig1}
\end{figure}

\begin{figure}[b]
\centering
 \includegraphics[width=2.9in]{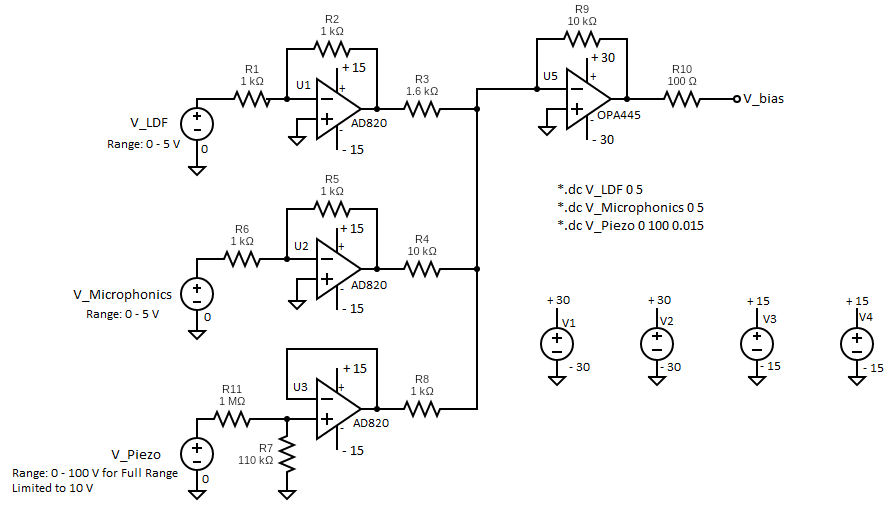}
\caption{Varactor Bias Summation Circuit}
\label {fig2}
\end{figure}

\section{Cavity Emulator Design}
\par
The crystal filter circuit section with the varactor diodes is shown in Fig. 1. The bias input is provided from the output of a signal summing circuit shown in Fig. 2. The inputs of the summing circuit are weighted in
the proportion shown in Table 1. The negative scale factor for the piezo input is for feedback control. The output bias voltage can be expressed in terms of its inputs as
\begin{equation}   
V_{BIAS} = V_{Micro} + 6.25 V_{LFD} - V_{Piezo}
\end{equation}

\begin{table}[!h]
\caption{Simulator Inputs}
\label{tab:sample}
\vspace{10pt}
\centering
\begin{tabular}{|c|c|c|}
\hline
Input & Range(V) & Scaling\\
\hline
Microphonics & 0 - 5 & 1\\
\hline
LFD & 0 - 5 & 6.25\\
\hline
Piezo & 0 - 30 & -1\\
\hline
\end{tabular}
\end{table}
The full schematic of the cavity emulator is shown in Fig.3.
\begin{figure*}[ht]
\centering
 \includegraphics[width=170mm]{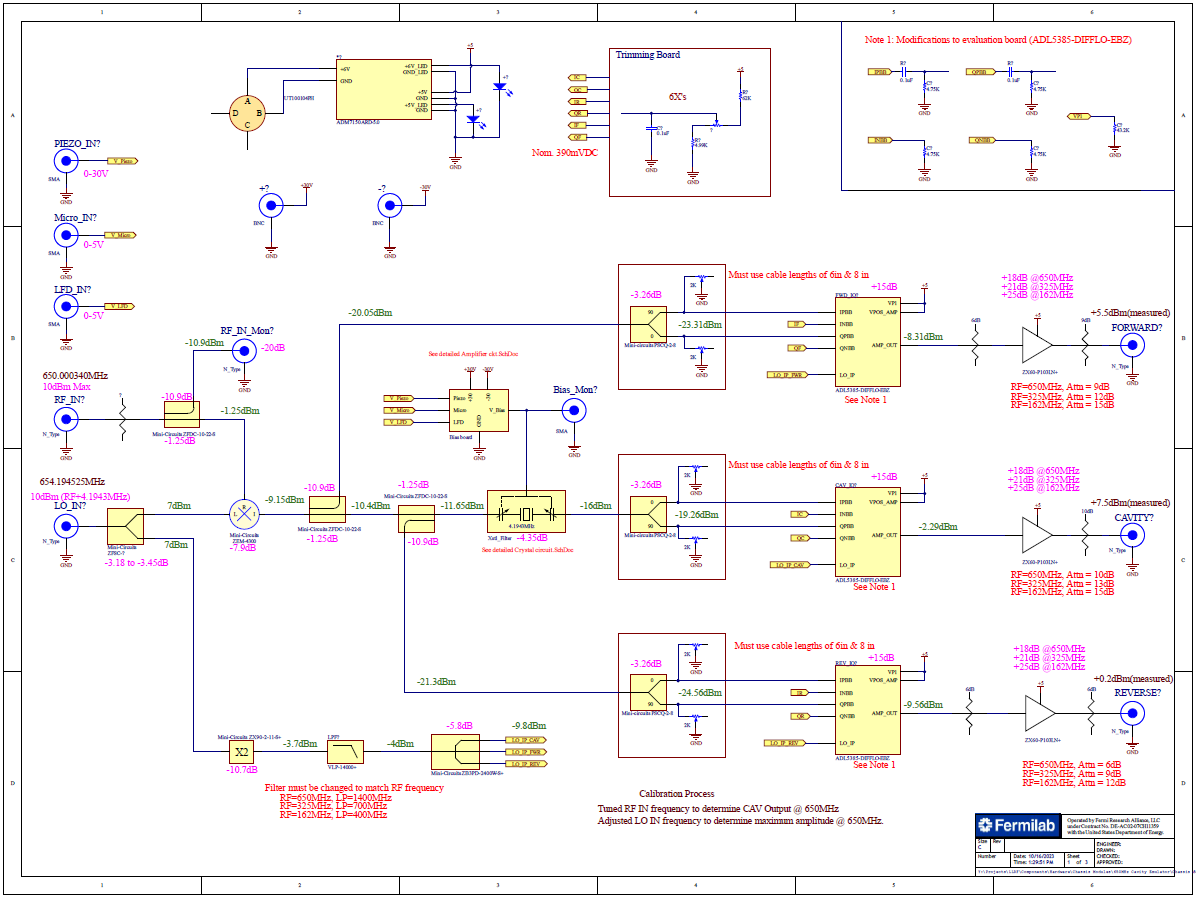}
\caption{Cavity Emulator Schematic}
\label {fig3}
\end{figure*}
\par 
The piezo input can reach a maximum of 100V which is seen as a maximum input of 10V due to the voltage divider
by a factor of ten at the piezo input. The gain in the summing stage is also 10 which provides an effective loop gain of one without using power supplies greater than 30V. For any piezo input greater than 30V the
output circuit will saturate. The numbers chosen here for the relative gains are not intended to provide any quantitative measure of the microphonics and LFD effects, but rather to provide a qualitative and relative
effectiveness of various control  schemes to suppress the amplitude of the resonances in the cavity detuning spectrum.
\par
The cavity emulator resonant frequency detuning range with the full bias  voltage of 30V is about 160 Hz as shown in Fig. 4. The detuning as a function of bias voltage is non-linear as seen in Fig. 6.
\begin{figure}[!b]
\centering
 \includegraphics[width=3.0in]{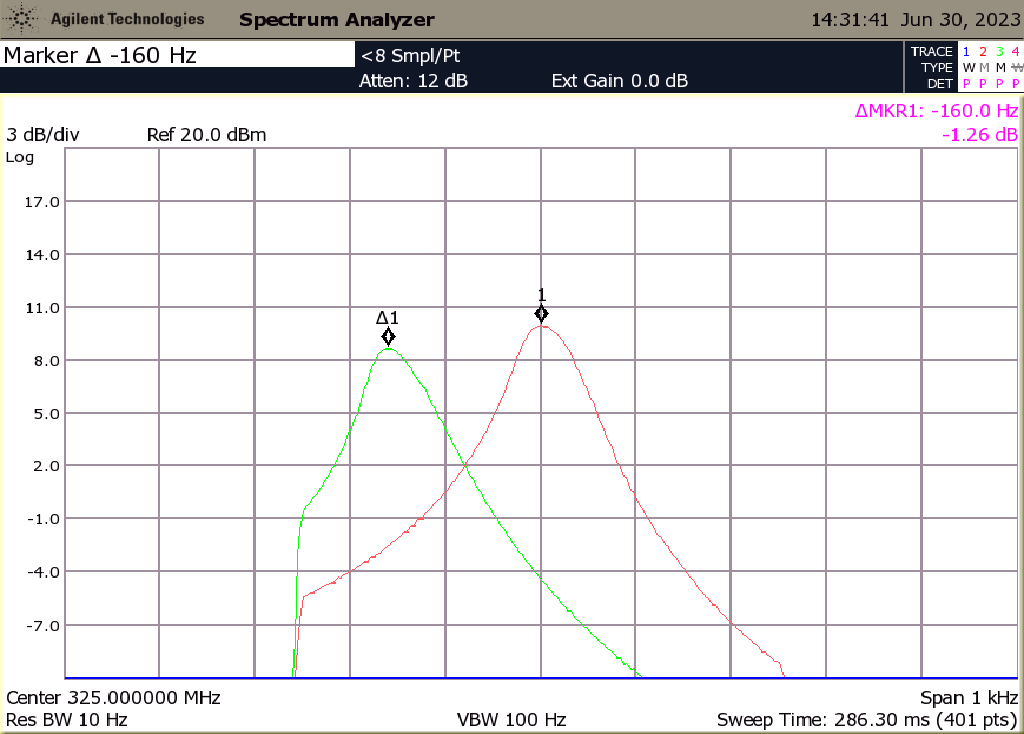}
\caption{Cavity emulator detuning range}
\label {fig4}
\end{figure}

\begin{figure}[!t]
\centering
 \includegraphics[width=3.0in]{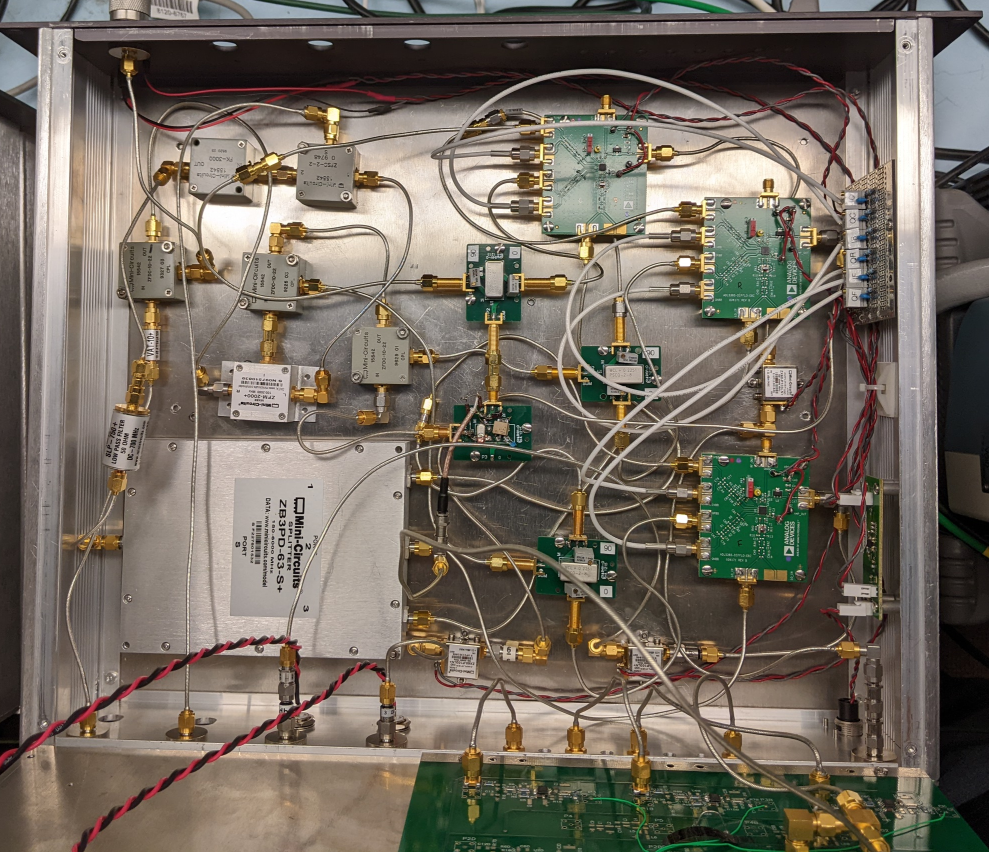}
\caption{Cavity Emulator}
\label {fig5}
\end{figure}

\section{Emulator Testing}
\par
A full LLRF hardware setup was assembled to test the emulator in closed loop. A schematic of the test setup is shown in Fig. 8. The hardware in the loop consists of a LLRF controller, a Resonance controller,
a master oscillator for the RF, clock and LO frequency generation, up and down converters for the RF and the cavity emulator. A picture of the setup is shown in Fig. 9.
\subsection{Open loop testing}
\par
A 20 Hz sinewave test waveform representing the cavity detuning was sent from a signal generator to the microphonics input and the resultant detuning is displayed in the 
Labview interface for the LLRF controller.The detuning data from the LLRF controller is transmitted over the fiber conection to the resonance controller. The non-linearity of the cavity detuning with bias voltage can be seen in the
distorted waveform seen for the cavity detuning in Fig.7.
\begin{figure}[!b]
\centering
 \includegraphics[width=3.0in]{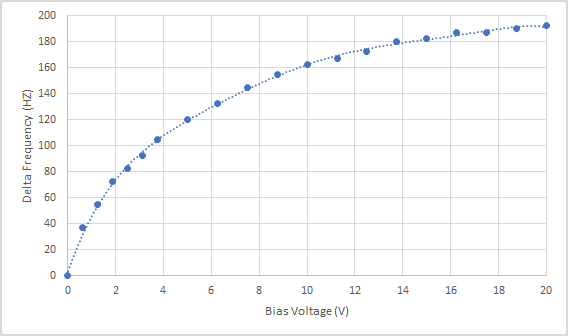}
\caption{Detuning vs. Bias Voltage}
\label {fig6}
\end{figure}
\begin{figure}[!t]
\centering
 \includegraphics[width=3.0in,height=2.0in]{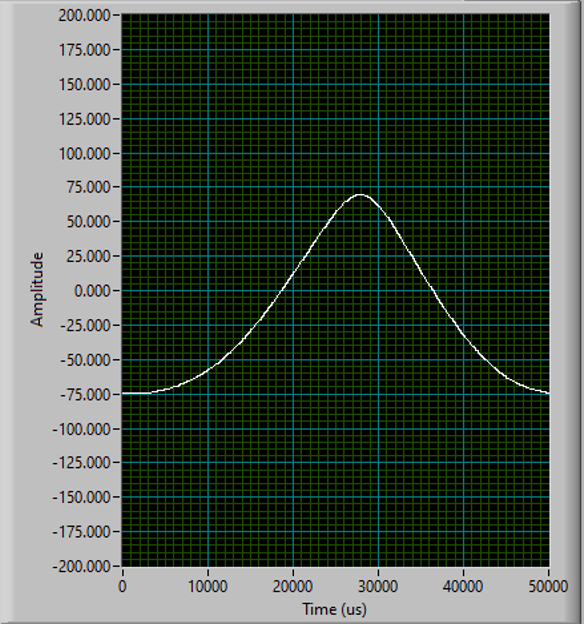}
\caption{Open Loop Test}
\label {fig7}
\end{figure}
\begin{figure}[!b]
\centering
 \includegraphics[width=3.0in]{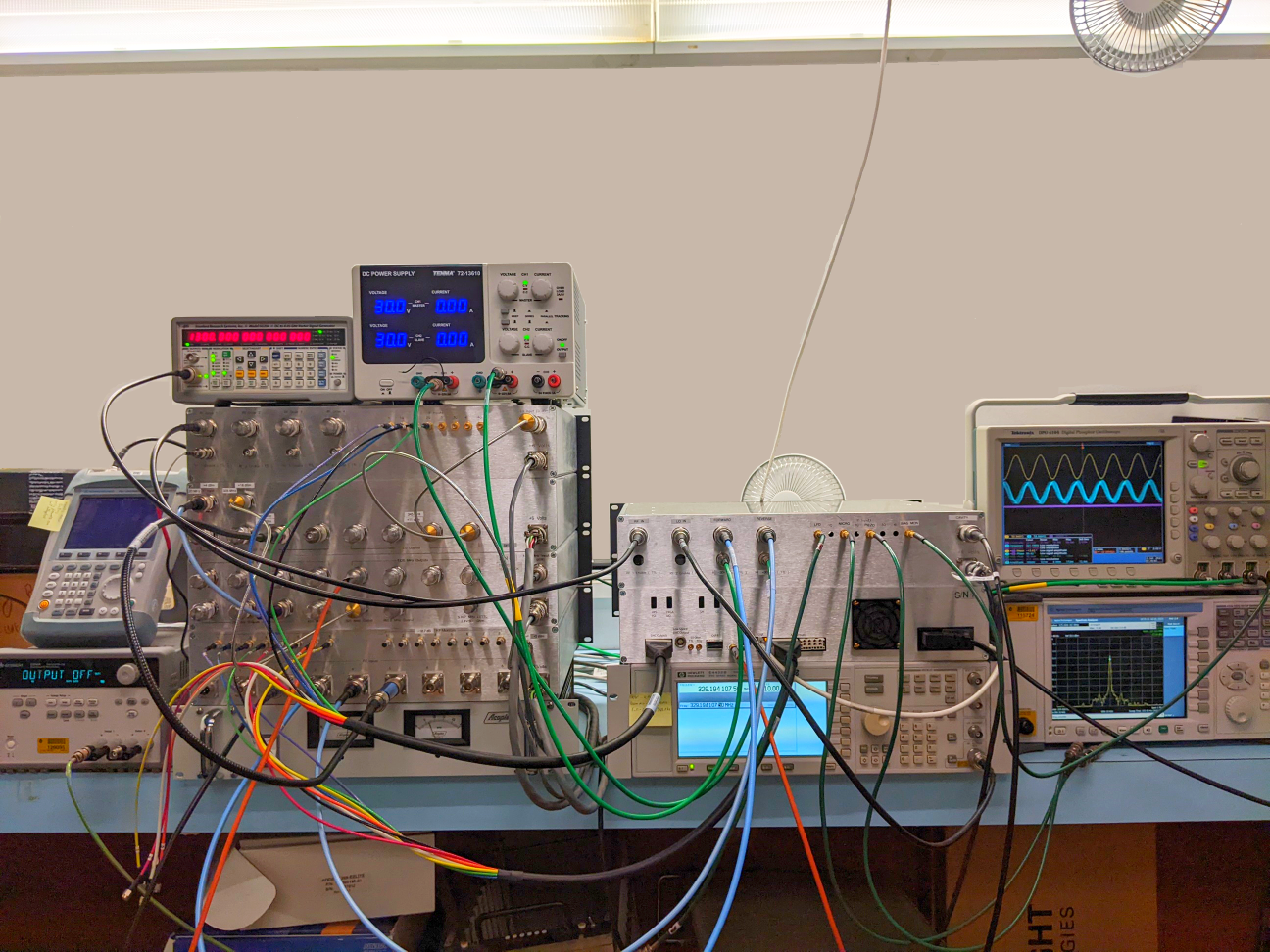}
\caption{Test Setup}
\label {fig8}
\end{figure}

\begin{figure*}[!t]
\centering
\includegraphics[width=160mm]{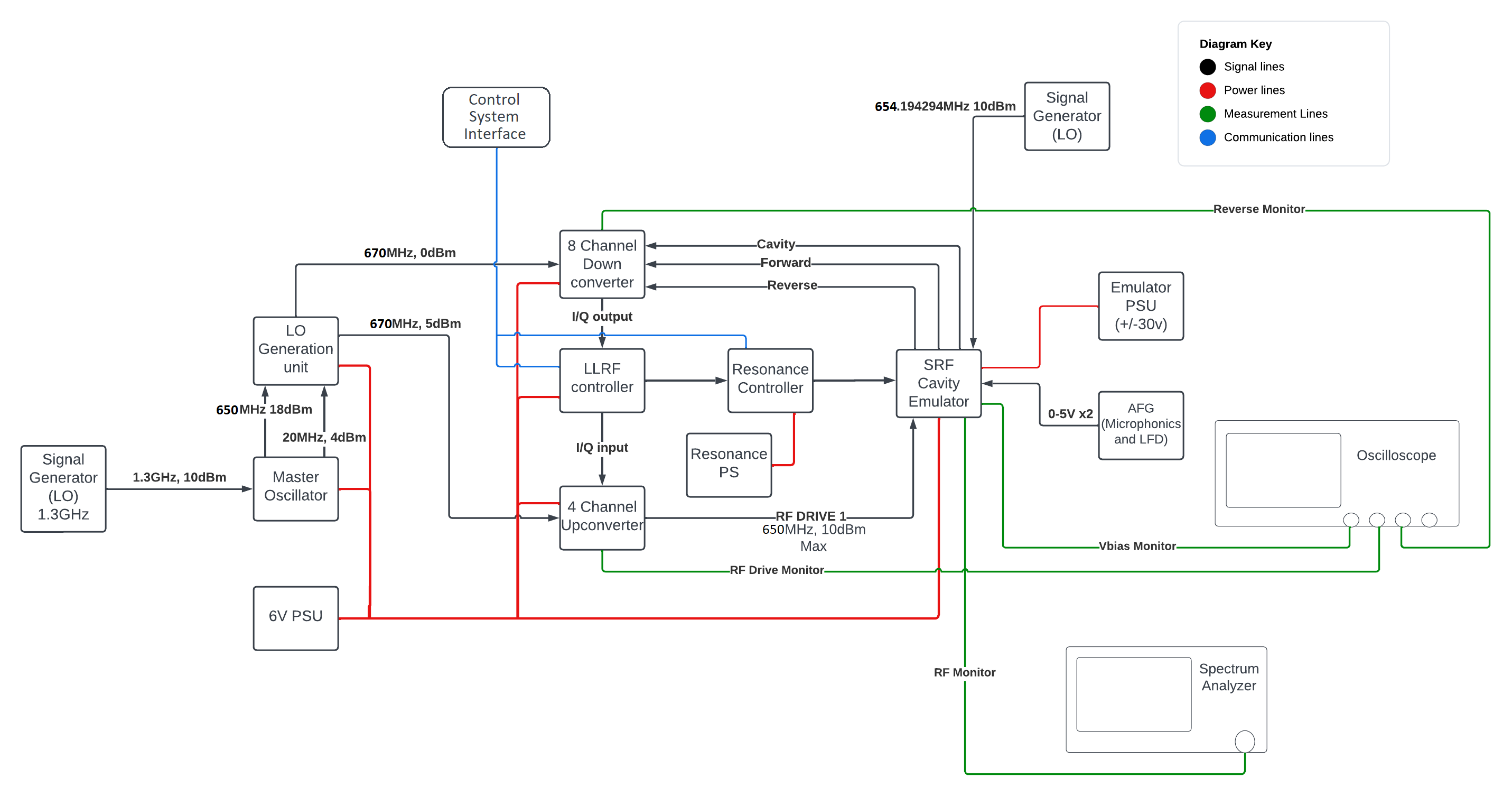}
\caption{Cavity Emulator test configuration with LLRF system}
\label {fig. 9}
\end{figure*}

\begin{figure}[!h]
\centering
 \includegraphics[width=2.8in]{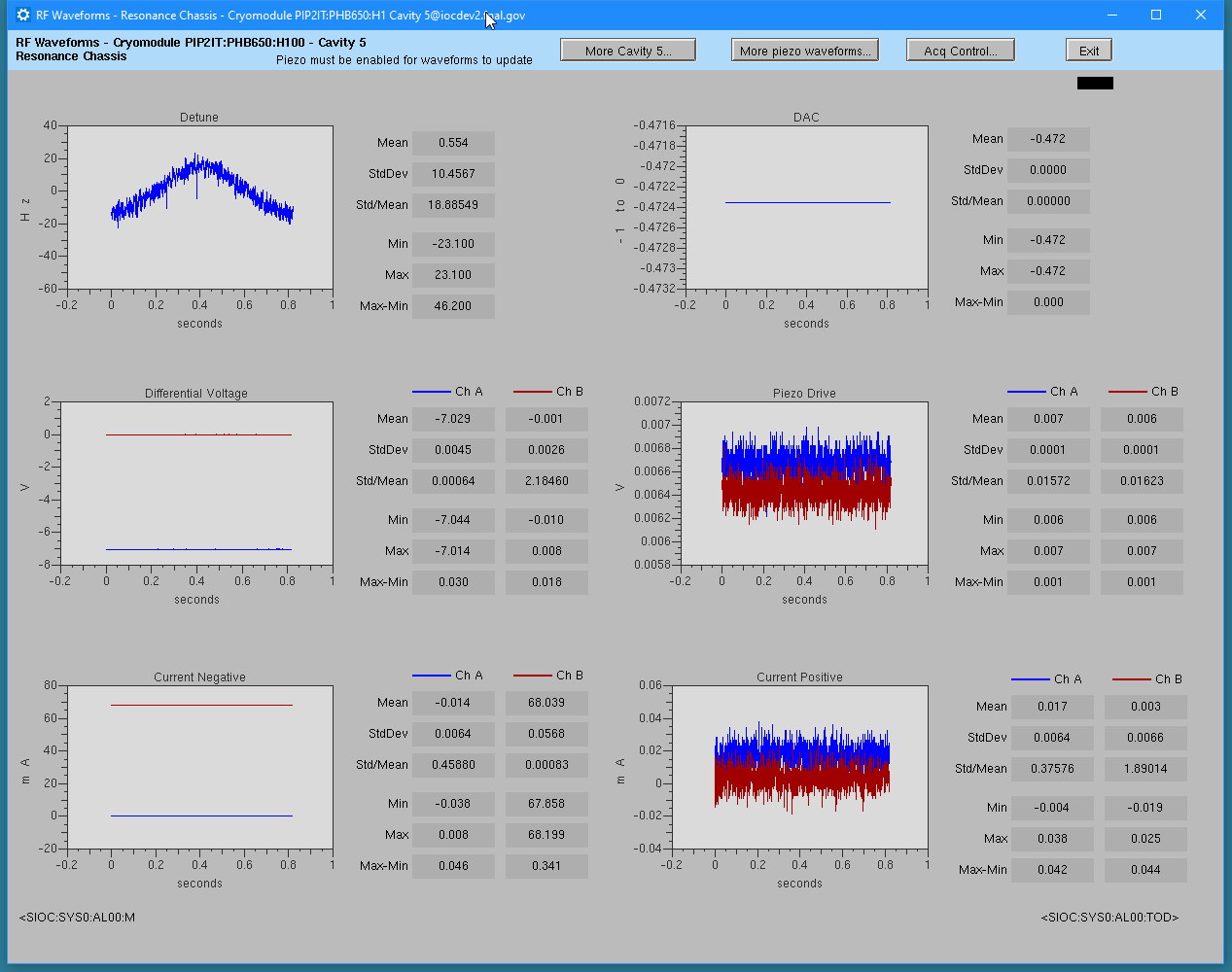}
\caption{Detuning with feedback disabled}
\label {fig10}
\end{figure}

\begin{figure}[!h]
\centering
 \includegraphics[width=2.8in]{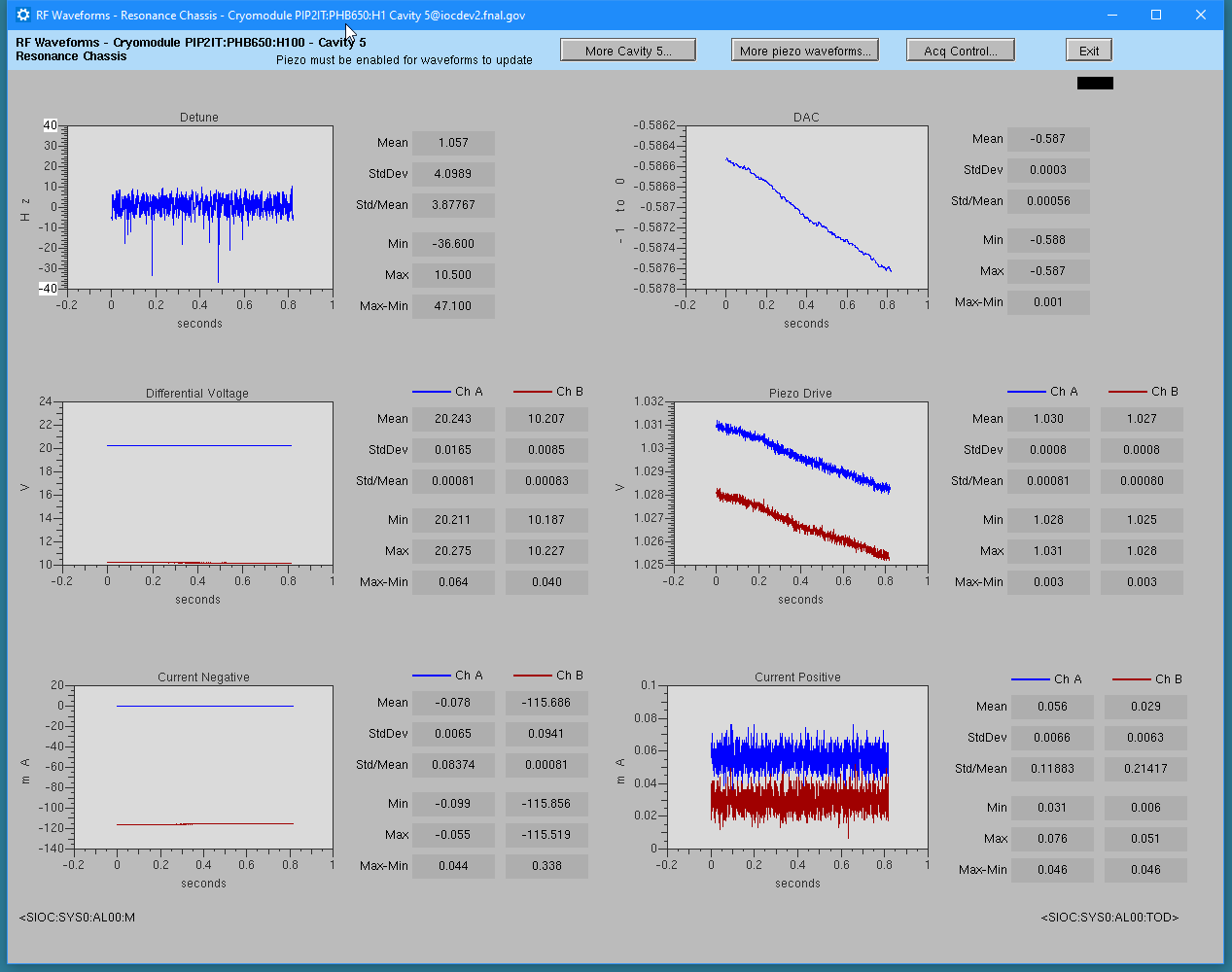}
\caption{Detuning with feedback enabled}
\label {fig11}
\end{figure}
 
\begin{figure}[!h]
\centering
 \includegraphics[width=2.8in, height = 2.2in]{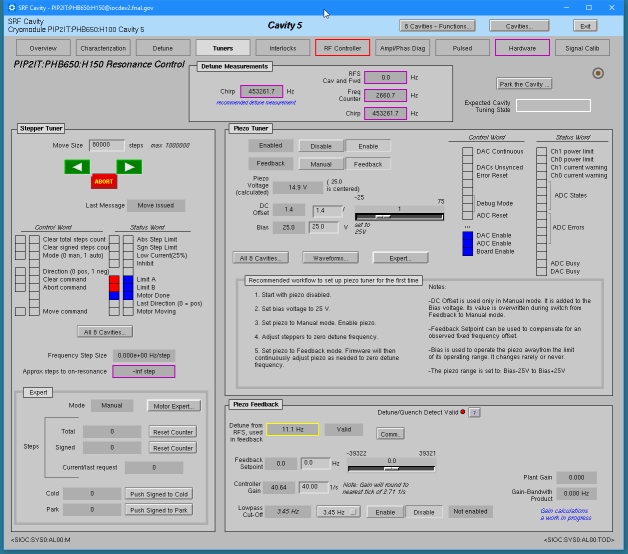}
\caption{Tuner Controls Settings}
\label {fig12}
\end{figure}

\subsection {Closed Loop Testing with Resonance Controller}
\par
The LLRF controller computes the cavity detuning from the cavity and forward power signals from the cavity emulator. The detuning data
is transmitted over a fiber channel to the resonance contoller. The processing of the detune data in the resonance controller is shown in Fig. 13. The piezo tuner drive is connected to the piezo input of the
cavity emulator.

\begin{figure*}[!t]
\centering
 \includegraphics[width=160mm, height=160mm]{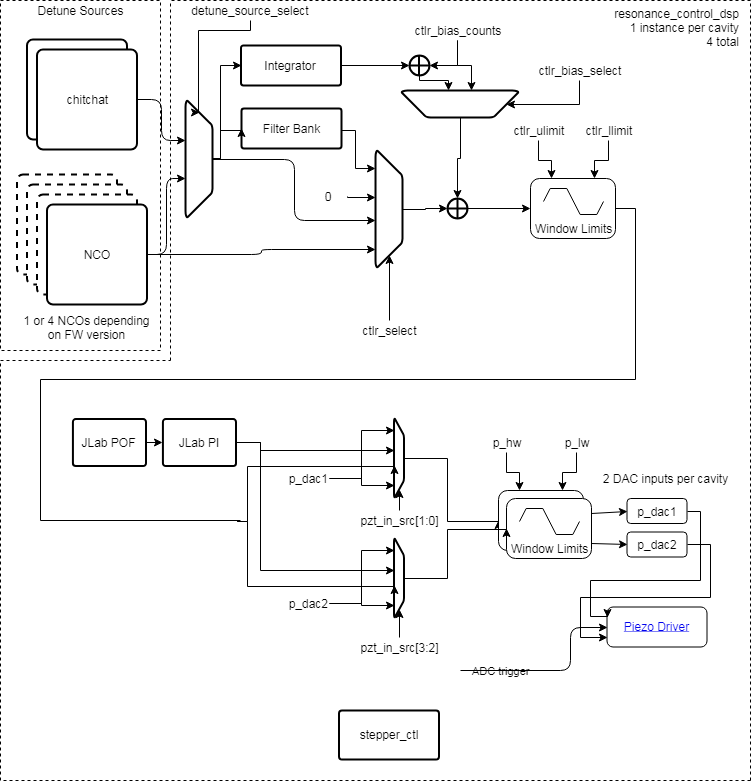}
\caption{Detuning Feedback Controller}
\label {fig13}
\end{figure*}

\par
The microphonics disturbance input can be driven by a signal generator driving the frequencies of interest. Cavity microphonics detuning data can be used to drive an arbitrary waveform generator to 
provide a disturbance input. The control loop in the resonace controller consists mainly of an integrator with a bandwidth of about 1.5 Hz that compensates for slow drifts in the cavity resonant frequency.
A filter bank with filters at different frequencies of interest with tunable gain and phase can be used to target specific microphonics and mechanical resonances of interest. 

\par
In order to test the slow drift integrator in the loop, a 1 Hz sinewave input was used to drive the disturbance input. The cavity detuning with the feedback off shows the effect of the disturbance as seen in Fig. 11.
When feedback is enabled the integrator can be seen to reduce the detuning frequency excursions as shown in Fig.12.
\par
The piezo voltage range input to the cavity emulator is 0-30V whereas the piezo drive output of the resonance controller is in the 0-100V with a differential drive of +/- 50 V. Since the summing circuit piezo input is single ended, this range is reduced to 0-50V. In order to keep the feedback loop from saturating the drive has to be less than 30 V. The microphonic disturbance input of a 1 Hz, 1Vpp sinusoidal input was given a DC offset of 0.5V. This resulted in a detuning range of about 40 Hz as seen in Fig. 11. Keeping the piezo drive bias at 20V provided sufficient headroom for the feedback loop to regulate without saturation as can be seen in Fig.12 which shows the piezo waveforms with feedback enabled. The frequency of 1 Hz was chosen for the disturbance because the feedback loop which consists of an integrator has a bandwidth in that range. When tuner loops with greater bandwidth are used, the circuit should be able to work in the full range of microphonics frequencies. The detuning measurement used a frequency PLL in the LLRF controller run in the cavity frequency tracking mode with a direct readout of the offset frequency from the nominal frequency. This approach was used to have better control of the test environment focusing on demonstrating the viability of this design. Detuning computed from cavity waveforms in GDR mode should work equally well. Microphonics data taken with real cavities can be scaled appropriately and driven by an arbitrary waveform generator to drive the disturbance input which will enable the tuner feedback algorithms to be evaluated for their effectiveness in suppressing the microphonics disturbances with real cavities.

\section{Summary}
\par
A crystal oscillator based analog cavity emulator was modified with reverse biased varactor diodes to have a tunable resonant frequency range of about 160 Hz. The resonant frequency shift with bias voltage
is used to emulate the effects of external microphonic disturbances or Lorentz force detuning. An analog summing circuit configuration with two inputs for disturbances and a third input for the correction voltage from the 
piezo controller was used to provide a suitable test bed to check the operation of piezo tuner closed loop operation in a feedback configuration. The emulator was tested with a complete LLRF system using a signal generator
to provide a microphonics disturbance. The piezo feedback loop was then turned on and the tuner was shown to significantly reduce the amplitude of the detuning oscillation. The use of a tunable crystal oscillator to
effectively emulate the impact of microphonic and other disturbances increases its utility in testing LLRF systems in the absence of superconducting  cavities or cryomodules.  This allows various features of the LLRF system
to be tested before cold cavities are available.


\begin{thebibliography}{9}   

\bibitem{}S. Ahmed, B. Chase, P. Varghese, S. Begum, “SRF Cavity Emulator for PIP-II LLRF Lab and Field Testing”, LLRF2022, Brugg-Windisch, Switzerland, October 2022.
\bibitem{}Feng Qiu, “Real-time cavity simulator-based low-level radio-frequency test bench and applications for accelerators", Phys. Rev. Accel. Beams 21, 032003, March 2018
\bibitem{}C.W. Pond, D. Kemper, "How to Specify Crystal Filters", Piezo-Electric Devices and Exhibition Conference, 1992 
\bibitem{}Paul R. Gray, Peter Baer Galvin, "Analysis and Design of Analog Integrated Circuits", John Wiley and Sons Inc., New York, New York, USA, 2001 
\end{thebibliography}
\end{document}